\documentclass[aps,prl,twocolumn,showpacs,preprintnumbers,groupedaddress]{revtex4}
\usepackage{graphicx}

\begin{document}
 
\draft

\renewcommand{\thefootnote}{\alph{footnote}}
\title{Lasing from a single quantum wire}

\author{
Yuhei Hayamizu\footnote[0]
{$^{a)}$Electronic mail: haya@issp.u-tokyo.ac.jp}$^{a)}$, 
Masahiro Yoshita,
Shinichi Watanabe,
and Hidefumi Akiyama
}

\address{
Institute for Solid State Physics, University of Tokyo,
5-1-5, Kashiwanoha, Kashiwa, Chiba, 277-8581, Japan}

\author{
Loren N. Pfeiffer
and Ken W. West
}
\address{
Bell Laboratories, Lucent Technologies, 600 Mountain Avenue, Murray Hill, NJ 07974, USA
}

\date{\today}

\begin{abstract}
A laser with an active volume consisting of only a single quantum wire in 
the 1-dimensional (1-D) ground state is demonstrated. The single wire is 
formed quantum-mechanically at the T-intersection of a 14 nm 
Al$_{0.07}$Ga$_{0.93}$As quantum well and a 6 nm GaAs quantum well, and is 
embedded in a 1-D single-mode optical waveguide. We observe single-mode 
lasing from the quantum wire ground state by optical pumping. The laser 
operates from 5 to 60 K, and has a low threshold pumping power of 5 mW at 5 
K.
\end{abstract}

\pacs{(78.67.Lt, 78.45.+h, 73.21.Hb, 42.55.Px.)}

\maketitle

\narrowtext


A quantum wire laser is a novel semiconductor laser based on a 1-dimensional 
(1-D) active region such that the electron and hole carriers are allowed to 
move only in that one direction. In contrast, more familiar semiconductor 
lasers are 3-D double-hetero-structure lasers and 2-D quantum well lasers. 
Because the 1-D density of states (DOS) becomes more enhanced at the bottom 
of its band edge than the 2-D or 3-D DOS, a quantum wire laser is expected 
to show improvement in lasing performance \cite{wei,arakawa,asada,arakawa2}. 

However, fabrication of a single quantum wire that lases from the lowest 
quantum state of the wire is difficult. Indeed all quantum wire lasers so 
far reported are either multiple-wire lasers \cite{weg,rubio,sirigu,kim,aki1,aki2}, which have many quantum 
wires in the active region, or they are wide single-wire lasers which lase 
only in their excited-subband states \cite{kapon,wata,pies}. The excited-state lasing is 
typically characterized by a lasing energy above the energy of ground-state 
spontaneous emission at low pump levels. Since excited subbands allow motion 
of carriers in directions perpendicular to the axis of the wire, such 
excited-state wire lasing is not expected to have 1-D characteristics. 

In this paper, we demonstrate a single quantum wire laser with only one 1-D 
subband, and observe stable single-mode lasing from this ground state 
subband of our quantum wire. 

The single quantum wire laser is fabricated by an advanced crystal growth 
method called cleaved-edge overgrowth with molecular beam epitaxy (MBE) 
\cite{pfei}, in which two MBE growth steps are separated by an \textit{in situ} wafer cleave. In the first MBE growth, we successively grew at 600 $^{o}$C on a non-doped (001) GaAs substrate, a 500 nm GaAs buffer layer, a 1.5 $\mu $m 
Al$_{0.5}$Ga$_{0.5}$As cladding layer, a 250 nm Al$_{0.35}$Ga$_{0.65}$As 
barrier, a 14 nm Al$_{0.07}$Ga$_{0.93}$As quantum well (stem well), a 250 nm 
Al$_{0.35}$Ga$_{0.65}$As barrier, 1.5 $\mu $m Al$_{0.5}$Ga$_{0.5}$As 
cladding layer, and a 6.5 $\mu $m GaAs cap layer. Then, we cleaved the wafer 
in the MBE chamber, and grew at 490 $^{o}$C on the newly exposed (110) edge, 
a 6 nm GaAs quantum well (arm well), a 10 nm Al$_{0.5}$Ga$_{0.5}$As barrier, 
a 111 nm Al$_{0.1}$Ga$_{0.1}$As core layer, a 960 nm Al$_{0.5}$Ga$_{0.5}$As 
cladding layer, and a 10 nm GaAs cap layer. After the growth of the arm 
well, we interrupted growth and annealed the GaAs surface for 10 minutes at 
600 $^{o}$C \cite{yoshi}.

\begin{figure}
\includegraphics[width=.45\textwidth]{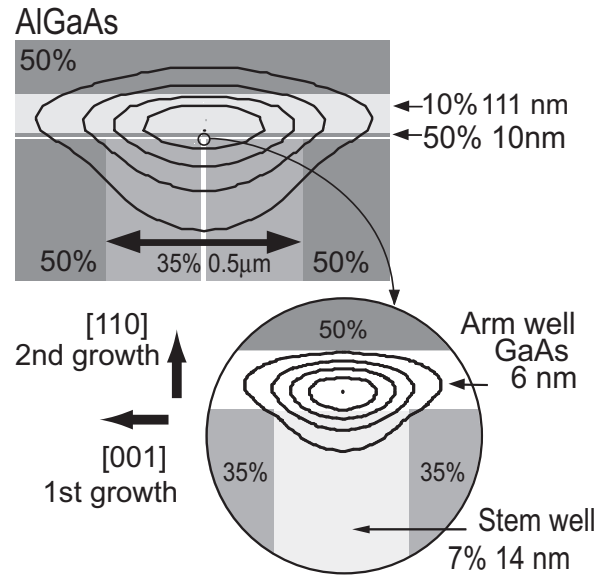}
\caption{
Schematic cross-sectional view of a single-wire laser structure. Percentages show Al-concentration \textit{x} in Al$_{x}$Ga$_{1 - x}$As. The constant probability ($\vert$\textit{$\phi $}$\vert$$^{2}$ = 0.2, 0.4, 0.6, 0.8 and 1.0) for photons and electrons in the device are drawn by top and bottom contour curves. 
}
\label{1}
\end{figure}

Figure 1 shows a schematic cross-sectional view of the single quantum wire 
laser structure. At a T-shaped intersection of a 14 nm-thick stem well and a 
6 nm-thick arm well, quantum-mechanical confinement of electrons forms a 
quantum wire (T-wire). Contour curves in the blow up of the T-wire in Fig. 1 
show the wave function of 1-D electrons. Previous experimental and 
theoretical studies \cite{aki1,szy} show that such a T-wire has no confined higher 
1-D electron subband, and is in the 1-D ground state quantum limit. A 
photoluminescence measurement shows that energies of the ground-state 
exciton in the T-wire, arm well and stem well are respectively 1.582, 1.602 
and 1.636 eV at 5K. The single quantum wire is embedded in a core of 
T-shaped optical waveguide (T-waveguide). Contour curves at the top part of 
Fig. 1 show optical field intensity, or optical mode (transverse mode), for 
lasing. The optical confinement factor $\Gamma $, defined as the portion of 
the optical intensity overlapped with the 14 nm $\times $ 6 nm region of the 
quantum wire, is optimized by choosing the thickness of the core layers in 
the T-waveguide on the basis of a finite-element-method calculation. In the 
present design, the value of $\Gamma $ is 5$\times $10$^{ - 4}$. 

A series of lasers each with 500 $\mu $m long optical cavities were cut from 
the wafer by (1$\bar{1}$0) cleavage, and the cavity-mirror surfaces were coated by 120 
nm- and 300 nm-thick gold layers with estimated reflectivities of 97 {\%}. 
Each laser was pumped optically with cw titanium-sapphire-laser light, 
mechanically chopped into a 1 {\%} duty ratio to minimize sample heating. 
The excitation light energy was 1.6455 eV. The pump laser light in a 
filament shape of about 1 $\mu $m width focused by a cylindrical lens and a 
0.5 numerical aperture objective lens was incident on the top (110) surface 
of the sample illuminating the 500 $\mu $m long T-wire uniformly. Laser 
emission was detected via the 120 nm-thick gold mirror which has an 
estimated transmissivity less than 0.1 {\%}.

\begin{figure}
\includegraphics[width=.45\textwidth]{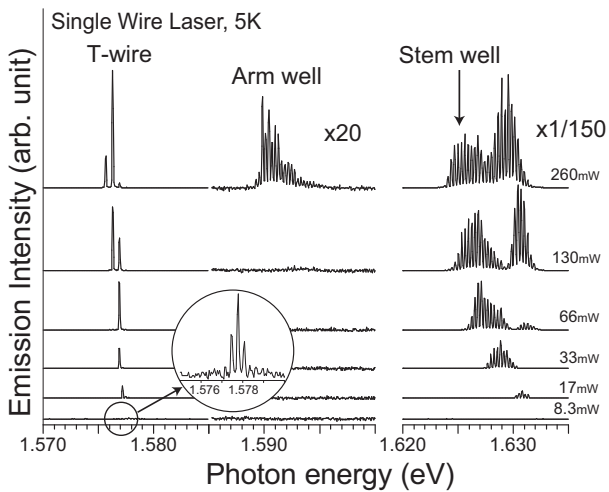}
\caption{
Lasing spectra of the single quantum wire laser at 5K for the various input powers of \textit{P}$_{in}$=8.3, 17, 33, 66, 130 and 130 mW, measured with spectral resolution of 0.2 meV. 
}
\label{2}
\end{figure}

Figure 2 shows typical laser emission spectra from a device at 5K for 
various excitation input powers $P_{in}$. At an input power $P_{in}$=8.3 mW, 
multi-mode laser emission of the T-wire is observed at 1.578 eV. It changes 
to single-mode at $P_{in}$=17 mW, and shows a slight red-shift with mode 
hopping as the input power increases. The energy stability of our quantum 
wire laser with increasing pumping power is remarkable. The total red-shift 
from laser threshold to pumping at 260 mW or $\times$ 50 threshold is only 1.5 meV. 
The details of this small shift with increasing pumping power are seen most 
clearly in Fig. 3. Note that for all pumping powers the lasing energies lie 
just below the ground-state exciton (1.582 eV) in the T-wire, proving that 
the lasing indeed occurs in the T-wire ground state. 

At higher energies, laser emission from the arm well and the stem well are 
also observed, as denoted in Fig. 2. Lasing by the arm well, shown with a 
magnified scale by a factor of 20, occurs only at very high pumping power 
$P_{in}$=260 mW. It cannot occur for lower powers, since photo-excited 
carriers in the arm well readily empty into the T-wire. At $P_{in}$=260 mW, 
however, the T-wire state saturates allowing the arm well to fill 
sufficiently to begin separately lasing. 

Lasing by the stem well is also observed for input powers above 
$P_{in}$=17 mW. Not surprisingly the emission is stronger than that of the 
T-wire, and is shown with a reduced scale by a factor of 1/150. In contrast 
to the case of the arm well, carrier flow from the stem well to the T-wire 
causes only a minor effect on carrier density in the stem well, since the 
excitation light generates carriers in the stem well over a large region 
extending many microns away from the T-wire. The stem well laser emission 
has multiple longitudinal modes. It shows a larger red-shift of 5 meV at 
$P_{in}$=260 mW. The stem well lasing can be regarded as a prototype of a 
gain-guided 2-D quantum well laser. 

It is interesting to compare the lasing of the T-wire with that of the stem 
well and the arm well, since it highlights in a single device intriguing 
differences in performance of a 1-D laser in comparison with 2-D lasers. 
First, the T-wire tends to lase in a single mode in contrast to multi-mode 
lasing of the arm and stem wells. This should be partly attributed to a 1-D 
optical waveguide, and to narrower gain spectra of the T-wire than those of 
the arm and stem wells. Second, the T-wire shows comparatively small 
red-shifts of 1.5 meV for increased input power $P_{in}$=260 mW, while the 
stem well shows larger red-shift of 5 meV. This reflects difference in 
many-body Coulomb interaction effects in a 1-D wire and a 2-D well \cite{aki3}. 
Third, the T-wire lasing has a lower threshold than the others. This last 
point is all the more remarkable since the T-wire laser has a small optical 
confinement factor $\Gamma $ = 5$\times $10$^{ - 4}$ in contrast to 
large $\Gamma $ = 0.02 in the stem well laser and $\Gamma $ = 0.01 in the 
arm well laser. Clearly the single T-wire laser has a high enough gain to 
compensate for the very low $\Gamma$ factor. These three features of 
the T-wire are also observed in single wire lasers we fabricated with 800 
$\mu $m-long cavities and also in multiple wire lasers with 20 wires \cite{aki2}.

\begin{figure}
\includegraphics[width=.45\textwidth]{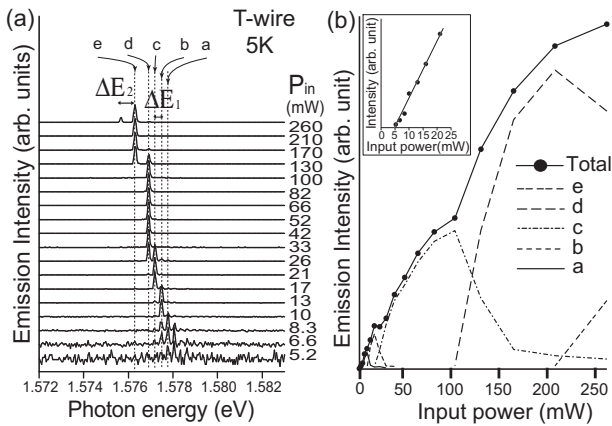}
\caption{
(a) Magnified lasing spectra of the single T-wire at 5 K, 
which are plotted in normalized scale for the various input powers 
\textit{P}$_{in}$ up to 260 mW. (b) Laser emission 
intensity of the T-wire at 5K against input powers up to 260 mW. The inset 
shows magnified plots near lasing threshold of 5 mW.
}
\label{3}
\end{figure}

Figure 3 (a) shows lasing spectra of the T-wire more in detail for the 
various input powers up to $P_{in}$=260 mW. Though the width of each emission 
line is limited by the spectral resolution 0.2 meV of the spectrometer, the 
results reveal interesting details of longitudinal modes in the T-wire 
lasing, where modes \textbf{a} - \textbf{e} denote respective longitudinal 
lasing modes. Multi-mode lasing with several longitudinal modes centered at 
\textbf{a} starts at input power of $P_{in}$= 5 mW. Transition from 
multi-mode lasing to single-mode lasing at \textbf{b} occurs at low input 
power of $P_{in}$=10 mW. Single mode lasing at such low input power is one of 
the remarkable characteristics of the T-wire laser. For increased input 
power, we observe mode hopping from \textbf{b} to \textbf{c}, to \textbf{d}, 
and to \textbf{e} with accompanying red-shifts. Below input power 
$P_{in}$=130 mW, lasing-mode separation, or mode hopping distance, is $\Delta 
E_{1}$=0.30 meV. This $\Delta E_{1}$ corresponds to the mode separation of 
a Fabry-Perot resonator with 500 $\mu $m cavity length, when the effective 
refractive index is 4.1. Above $P_{in}$=130 mW, however, it becomes $\Delta 
E_{2}$=0.60 meV. The reason of the doubled mode separation at high pumping 
is not understood. 

Figure 3 (b) plots laser emission intensity of the T-wire at 5K as a 
function of input power. The inset shows magnified plots near threshold, 
which gives the value of the threshold input power at 5 mW. 

For higher input powers above $P_{in}$=20 mW, laser emission intensities of 
the T-wire plotted by black dots and denoted as Total in Fig. 3 (b) 
increases showing irregular oscillations and dips. We also plot intensities 
of respective longitudinal modes \textbf{a} - \textbf{e} in T-wire lasing. 
It turns out that the dips in the total intensity occur exactly at those 
input powers where the laser switches from one lasing mode to the next. The 
reason for this interesting effect is also not at present understood.

\begin{figure}
\includegraphics[width=.45\textwidth]{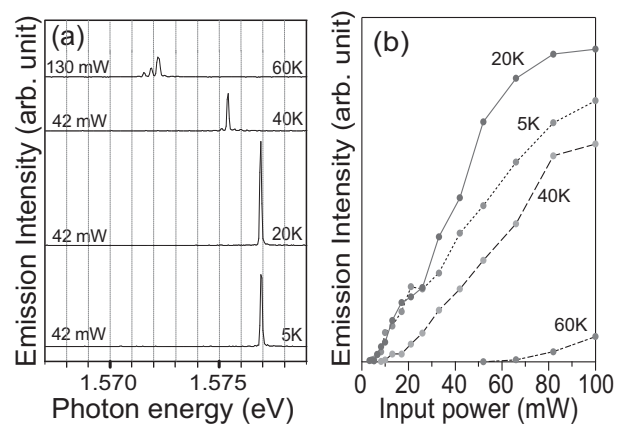}
\caption{
(a) Lasing spectra of the single T-wire at 5, 20, 40 and 
60K. (b) Laser emission intensity plots for the single T-wire at 5, 20, 40 
and 60K against input powers up to 100 mW.
}
\label{4}
\end{figure}

Figure 4 (a) shows the laser emission spectra of the T-wire at 5, 20, 40 and 
60K for input powers of 42 mW and 130 mW. Single-mode lasing is observed up 
to 40K, while only multi-mode lasing occurs up to 60K. With the increase in 
temperature, the spectral peaks are shifted toward lower energies as 
expected due to the shrinkage in the GaAs band-gap energy.

Figure 4 (b) plots laser emission intensity of T-wire at 5, 20, 40 and 60 K 
against input powers up to 100 mW. Notice that the lasing intensities at 20 
K are larger than those at 5K. This is caused by enhanced carrier diffusion 
in the stem well and the arm well which provides more carriers to the T-wire 
at 20K than at 5K. At 40 K and 60 K, however, lasing intensities become 
smaller and the lasing threshold increases. This is probably because 
carriers in the T-wire are thermally activated to escape to distant regions 
of the arm well and the stem well. 

In summary, we fabricated a single quantum wire laser in the 1-D quantum 
limit. We observed lasing at the ground state of a 14 nm $\times $ 6 nm wire 
at 5 - 60 K via optical pumping. It shows a low threshold input power of 5 
mW at 5 K. 

One of the authors (H. A.) acknowledges the financial support from the 
Ministry of Education, Culture, Sports, Science and Technology, Japan.

\end{document}